\documentclass{nature}
\usepackage{amsmath,amssymb,bm}
\usepackage{graphicx}
\usepackage{epstopdf}
\usepackage{caption}
\usepackage{color}

\usepackage{tabularx}
\usepackage{longtable}
\usepackage{booktabs}
\usepackage{nicefrac}

\newcommand{\br}[1]{\left( #1 \right)}

\title{Surface induced electronic Berry curvature in Berry curvature free bulk materials} 

\author
{Dennis Wawrzik$^1$, Jorge I. Facio$^{1,2}$, Jeroen van den Brink$^{1,3}$\\
\\
\normalsize{$^{1}$Institute for Theoretical Solid State Physics, IFW Dresden, Helmholtzstr. 20, 01069 Dresden, Germany}\\
\normalsize{$^{2}$Centro Atomico Bariloche, Instituto de Nanociencia y Nanotecnologia (CNEA-CONICET) and Instituto Balseiro. Av. Bustillo 9500, Bariloche (8400), Argentina}\\
\normalsize{$^{3}$Institute for Theoretical Physics and Cluster of Excellence ct.qmat, TU Dresden, 01069 Dresden, Germany}\\
\\
}

\date{\today}

\begin{document} 


\baselineskip22pt


\maketitle 

\begin{abstract}
In recent years it has become clear that electronic Berry curvature (BC) is a key concept to understand and predict physical properties of crystalline materials.
A wealth of interesting Hall-type responses in charge, spin and heat transport are caused by the BC associated to electronic bands inside a solid:
anomalous Hall effects in magnetic materials\cite{Karplus1959,Chang2013}, and various nonlinear Hall\cite{PhysRevLett.115.216806,Moore2010,Ma2018} and Nernst effects\cite{Xiao2006,Yu2019} in non-magnetic systems that lack inversion symmetry.
However, for the largest class of known materials --non-magnetic ones with inversion symmetry-- electronic BC is strictly zero\cite{Xiao2010,Vanderbilt2018}.
Here we show that precisely for these bulk BC-free materials, a finite BC can emerge at their {\it surfaces} and interfaces.
This immediately activates certain surfaces in producing Hall-type transport responses.
We demonstrate this by first principles calculations of the BC at bismuth, mercury-telluride (HgTe) and rhodium surfaces of various symmetries, revealing the presence of a surface Berry curvature dipole and associated quantum nonlinear Hall effects at a number of these.
This opens up a plethora of materials to explore and harness the physical effects emerging from the electronic Berry curvature associated exclusively to their boundaries.
\end{abstract}

\pagebreak

\maketitle

Inside a solid the anomalous velocity of Bloch electrons caused by the electronic Berry curvature (BC) leads to a sideways deflection of currents, much as an external magnetic field\cite{Karplus1959,Xiao2010,Vanderbilt2018}. Such a current deflection caused by the intrinsic electronic structure of a material and its associated Hall-type response must be allowed by the symmetries of the system. It is well-known that an absence of magnetism, and consequently a presence of time-reversal symmetry in the electronic band structure, leads to {\it zero total} BC, integrated over Brillouin-zone momenta. This consequently forbids any {\it linear} Hall response.
But as long as at least lattice inversion symmetry is broken, a finite BC is associated to the electronic structure, which then can allow, e.g., a quantum {\it nonlinear} Hall effect where the observed transversal voltage is proportional to the applied current squared. This nonlinear response is related to the dipole moment of the Brillouin-zone integrated BC, the Berry curvature dipole (BCD). For a bulk BCD to be present in a system with bulk BC, the crystal must belong to certain point groups\cite{PhysRevLett.115.216806}.

However, for materials with {\it both} time-reversal invariance {\it and} inversion, the BC  vanishes identically at all momenta in the 3D Brillouin-zone. So from the BC point of view such materials seem to become uninteresting as their anomalous electron velocities now appear to vanish as well. 
We demonstrate that while this conclusion is valid for 3D bulk materials, at surfaces (and interfaces) of these the opposite holds: at a general Miller index surfaces Bloch-electrons {\it do} attain a finite anomalous velocity, {\it also} for materials with bulk inversion and  time-reversal symmetry. 
Indeed we find that as a general rule of thumb the surface BC only vanishes at unreconstructed low Miller index surfaces, whereas a finite surface BC is associated to any higher index surface, including miscut and vicinal ones with surface steps that break C$_2$-rotation around the surface normal. 
If there is in addition not more than one mirror plane containing the normal, a finite surface BCD forms. Very many elementary surfaces fulfil this symmetry requirement.

To  explicitly evaluate the surface Berry curvature (SBC) and the SBC dipole (SBCD) for real materials we start from symmetry-conserving maximally-projected Wannier functions from density-functional calculations using the Full-Potential-Local-Orbital code, FPLO\cite{PhysRevB.59.1743,koepernik2021symmetry}, in order to construct bulk Hamiltonians. We subsequently obtain the surface Green's function for a semi-infinite slab of material using an iterative construction (see Methods).  After extracting the surface Green's function from the bulk Hamiltonian we evaluate its associated SBC. The numerically efficiency of calculating the Green's function for a semi-infinite system sets this method apart from a direct real-space slab approach, the numerical cost of which strongly increases with slab thickness and which would in addition require disentangling the BC contribution of different layers in the finite slab. 

\begin{figure}
  \centering
  \includegraphics[width=.5\columnwidth]{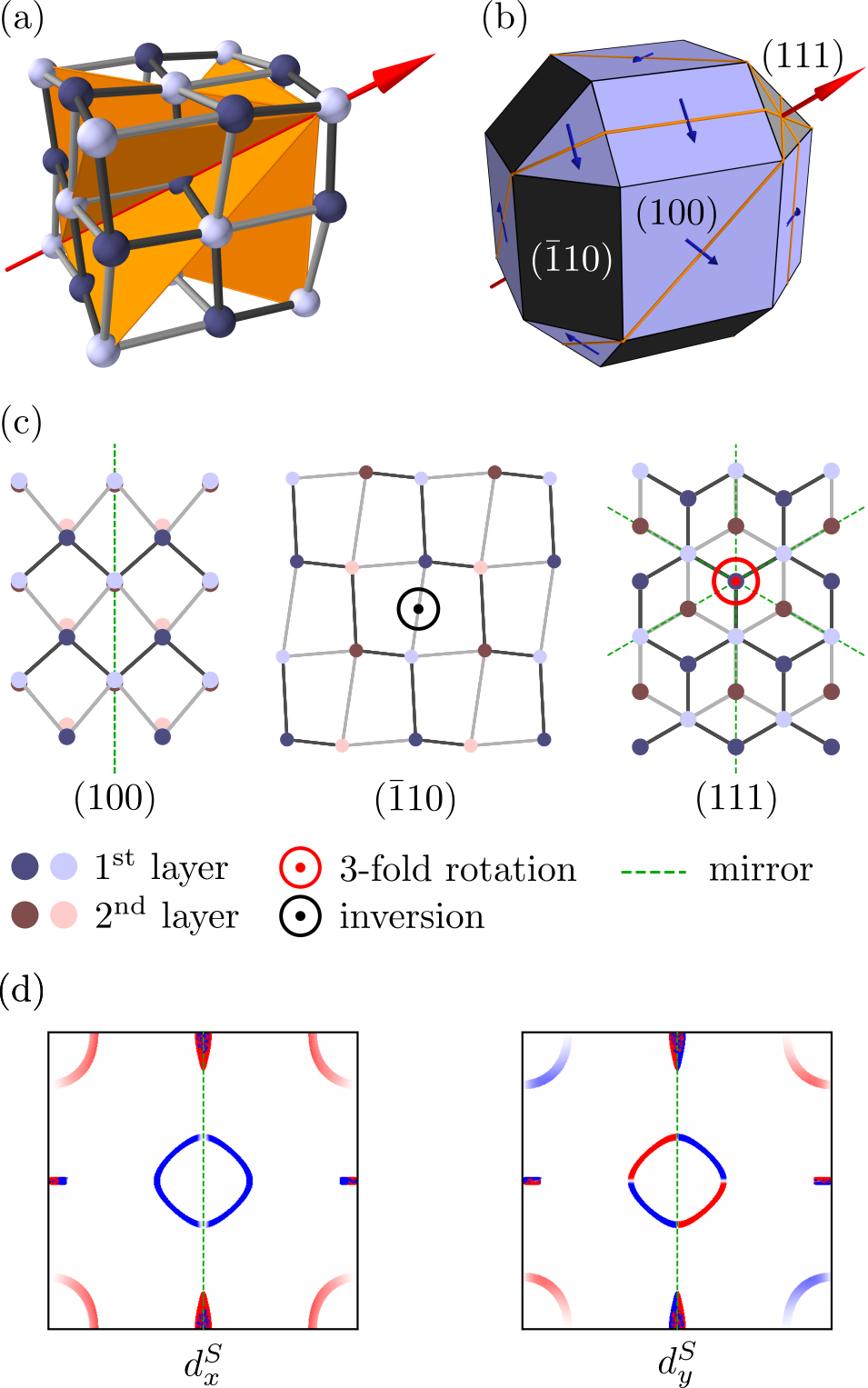}
	\caption{\textbf{Berry curvature at elemental bismuth surfaces. } (a) Lattice structure and symmetries of bismuth. Mirror planes are marked in orange and the 3-fold rotoinversion is denoted by the red arrow. (b) Surface Berry curvature (SBC) of different facets of a Bi crystal. Blue facets have a SBC dipole perpendicular to the mirror line while there is no SBC on black facets. The grey surface possesses SBC but no SBC dipole. (c) Symmetries of certain Bi surfaces. (d) SBC dipole density $\vec{d}^S$ of the (100) surface where the mirror symmetry causes vanishing $\vec{D}^S_y$ and finite $\vec{D}^S_x$. An electric field parallel to the SBC dipole, $\vec{E}\parallel \vec{D}^S$, results in a nonlinear perpendicular in-plane current $\vec{j}\perp\vec{D}^S$ along the mirror line.}
  \label{Fig:Bi_sym}
\end{figure}

\begin{figure}
  \centering
  \includegraphics[width=.75\columnwidth]{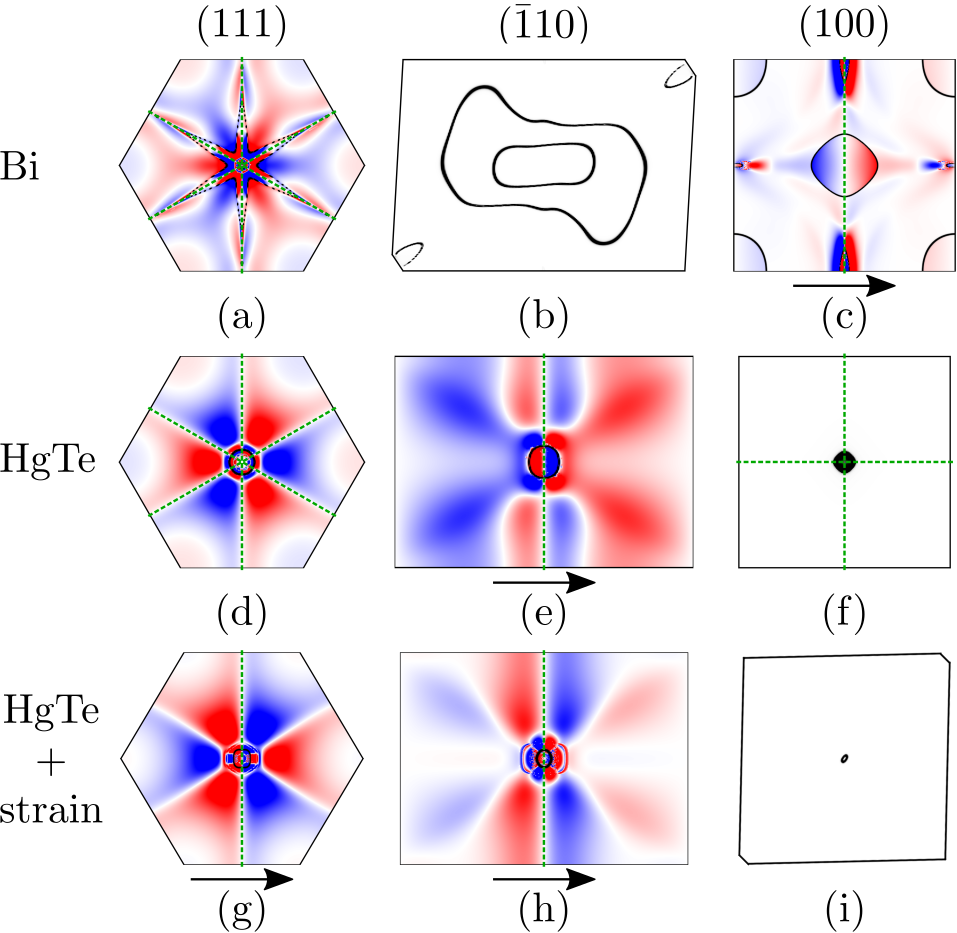}
	\caption{\textbf{Electronic surface Berry curvature.} From left to right, data corresponds to (111),  ($\bar{1}$10) and (100) surfaces for (a)-(c) Bismuth, (d)-(f) HgTe, and (g)-(i) strained HgTe. The Fermi surface is marked in black and dashed green lines denote mirror symmetries. If a surface has a SBCD its direction is marked by the black arrow.} 
  \label{Fig:BC_Bi_HgTe}
\end{figure}

We demonstrate the principles and method outlined above by considering three different surfaces of inversion and time-reversal invariant elemental bismuth, a strictly BC-free bulk crystal. The effects of (surface) symmetry can be nicely illustrated on Bi crystals, which have a 3-fold rotoinversion axis (cf.\ Fig.\ \ref{Fig:Bi_sym}(a-b)) that cancels all bulk BC. The ($\bar{1}$10) surface\cite{footnote1} of Bi  has a two-dimensional center of inversion (which is equivalent to a C$_2$-rotation symmetry around the surface normal) and thus no SBC, whereas at the (111) and (100)  surfaces inversion is broken, see Fig.\  \ref{Fig:Bi_sym}(c). Consequently Bi (111) and (100) have a finite SBC associated to them, as the results of a direct computation shown in Fig.\ \ref{Fig:BC_Bi_HgTe}(a-c) confirm. Since the bulk inversion relates top and bottom surface, opposite surfaces will have opposite SBC. Consequently the {\it total} BC for an inversion symmetric piece of bulk crystal including its surfaces, vanishes again. We note that for Bi (100) in Fig.~\ref{Fig:BC_Bi_HgTe}(c) the  projection of the bulk states gives a large contribution to the BC close to  $\vec{k}_\parallel=(0,\pi)$ while the states around $\Gamma$ and $\vec{k}_\parallel=(\pi,\pi)$ are clearly separated from the bulk.

From the symmetry point of view the Bi (100) surface has only a single mirror line, as depicted in Fig.\ \ref{Fig:Bi_sym}(c). Thus it allows for a non-zero SBCD that in principle is accessible experimentally: an electric field parallel to the SBCD will result in a nonlinear current perpendicular to it, flowing along the mirror line, see Fig.~\ref{Fig:BC_Bi_HgTe}(c). The SBCD on other crystal facets with a single mirror line will also give rise to a current in direction of the 3-fold rotation axis, as depicted in Fig.\ \ref{Fig:Bi_sym}(b). To quantitatively determine the BCD associated to the Bi (100) surface states we evaluate the SBC dipole ${\vec D}^S$  for a k-mesh with $512^2$ k-points and an energy window $\Delta\mu=0.04\,\text{eV}$, see Fig.\ \ref{Fig:Bi_sym}(d). This yields $D^S_x = 0.03\,\text{\AA}$ for the state at $\vec{k}_\parallel=(\pi,\pi)$ and $D^S_x = -0.22\,\text{\AA}$ at $\Gamma$. Of course deeper layers will also contribute to the total BC dipole associated to this surface, but it is interesting to note that the top layer by itself   already has a BCD that is of the same order as the BCD of monolayer WTe$_2$~\cite{You2018}.

The binary compound mercury-telluride, HgTe, has gained much interest as the basis material for topologically insulating phases\cite{Koenig07,PhysRevLett.106.126803}. Its zincblende structure breaks inversion allowing for a finite bulk BC but has too many symmetries for a bulk BC dipole. Similar to Bi we find a finite SBC on the (111) surface but no SBCD due to the 3-fold rotation and the three mirror lines, see Fig.~\ref{Fig:BC_Bi_HgTe}(d). In contrast, HgTe possesses two-fold rotation axes which is naturally preserved by surfaces perpendicular to it. Since a two-fold rotation is identical to inversion in 2D, the HgTe (100) surface has zero SBC (Fig.~\ref{Fig:BC_Bi_HgTe}(f)). On the other hand, the ($\bar{1}$10) surface only has a single mirror line yielding a clear SBCD (Fig.~\ref{Fig:BC_Bi_HgTe}(e)). This implies that the natural (110) cleaving plane\cite{Liu2015} of HgTe, has a SBCD associated to it. 

A feature setting three-dimensional bulk HgTe apart is it becoming a topological insulator when tensile strain is applied and a gap opens up. Straining HgTe along the xy-plane, the inversion on (100) and the mirror on ($\bar{1}$10) remain and a SBC is associated to the topological surface state on ($\bar{1}$10). Strain breaks the 3-fold rotation and several mirror planes and consequently the strained HgTe (111) surface, and in particular its topological surface state, now also acquires a finite SBCD, as shown in Fig.\ \ref{Fig:BC_Bi_HgTe}(g-i). 

\begin{figure} 
\centering 
  \includegraphics[width=.75\columnwidth]{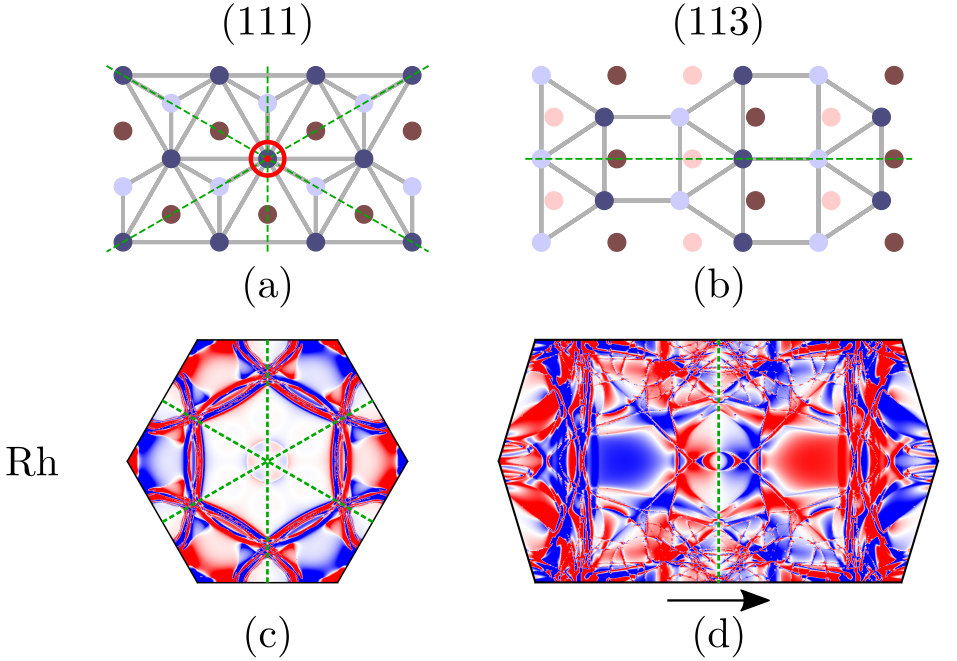}
	\caption{\textbf{Surface structure and Berry curvature of elemental fcc rhodium.} (a)-(b) (111) and (311) surface geometry and symmetries (c)-(d) surface Berry curvature (SBC) of (111) and (311) top layers. Whereas bulk Rh is Berry curvature free, Rh (111) surface supports a SBC and Rh (113) even a SBC dipole.  Legends as in Figs.~\ref{Fig:Bi_sym},\ref{Fig:BC_Bi_HgTe}.}
  \label{Fig:Rh_SBC}
\end{figure}

Similar considerations as the ones above for Bi and HgTe apply to other interfaces and materials. For example, a plethora of noble metal surfaces are experimentally available with electronic properties that that are important for catalysis~\cite{Li2020}. From our analysis we conclude that (111) surfaces of hcp and fcc noble metals (e.g., Au, Pt, Rh, Ir) have a finite SBC, as is illustrated by a direct calculation for elemental Rh in Fig.~\ref{Fig:Rh_SBC}. For such (111) surfaces the SBCD vanishes but it can become finite when reducing the symmetry, e.g. by applying uniaxial strains as shown above for HgTe (111). Since these noble metals are used as substrates of other 2D materials\cite{gao2021controllable}, e.g. graphene\cite{PhysRevLett.108.066804}, a proper use of the latter can be another interesting path to reduce the symmetry and induce a finite SBCD. Alternatively, higher Miller index surfaces can be considered.  In Fig.~\ref{Fig:Rh_SBC}(d) we show the BC associated to Rh (311),  which is a stepped surface with a symmetry reduced to a single mirror line, resulting in a SBCD. Similarly a SBCD is  also associated to, e.g., chemically relevant  Rh (210)~\cite{rebholz1992subsurface} and Rh (310)~\cite{ren1990study} surfaces. The rather complex Fermi surface of bulk Rh produces an intricate Rh (311) SBC distribution in Fig.~\ref{Fig:Rh_SBC}, certain detail of which depends on longer range-matrix elements (see Methods), but it  has a dstinct SBCD associated to it due to the presence of just single mirror line.

Having established the presence of SBCs and SBCDs at the boundaries of a very large class of materials, including the semimetal Bi, topological HgTe and the transition metal Rh, it should be very interesting to corroborate these theoretical findings by experimentally determining the momentum-resolved surface Berry curvature by dichroic angle-resolved photoelectron spectroscopy\cite{Schuler2020}, and to observe the emerging surface Berry curvature dipole that we predict here by the presence of a nonlinear Hall effect at relevant surfaces and interfaces.

\pagebreak

\section{Methods}
{\it Berry Curvature from Surface Green's function}  ---
Associated to the Hamiltonian matrix $H_{\bm k}$ of a 3D system is the imaginary-frequency Green's function $G(i\omega,{\bm k} ) = (i \omega - H_{\bm k})^{-1}$ and a Berry curvature of all occupied bands given by~\cite{ISHIKAWA1987523,PhysRevLett.113.136402}
\begin{equation}
  \Omega_c({\bm k})=\int_{-\infty}^{\infty}\frac{d\omega}{2 \pi} \text{tr}\br{g_c(i\omega,{\bm k})}
  \label{Eq:BC}
\end{equation} with
\begin{equation}
  g_c(i\omega,{\bm k} ) = \frac{\epsilon_{\mu \nu \rho c}}{3!}~G \br{\partial_\mu G^{-1}} G \br{\partial_\nu G^{-1}} G \br{\partial_\rho G^{-1}}
  \label{Eq:BC_Green}
\end{equation}
where $c \in \{x,y,z\}$ and $\mu,\nu,\rho \in \{\omega,x,y,z\}$.
Whereas evaluating this Green's function expression is not a very efficient method to calculate the BC for a 3D band structure, we rely on it to compute the SBC after extracting the surface Green's function from the bulk Hamiltonian.
The iterative algorithm in Ref.\ \cite{san85} is usually used to compute surface spectral functions of semi-infinite systems. The method can efficiently calculate the Green's functions  $G^S(i\omega,{\bm k} _\parallel)$ of the top, bottom, and central layers of a system of size $L$ for very large $L$
by doubling the slab size in every iteration. The corresponding tight-binding Hamiltonian $H(\vec{k}_\parallel)$ must be of the form
\begin{align}
    H = 
    \begin{pmatrix}
        H_0         & H_t         & ~      & ~           & ~   \\
        H_t^\dagger & H_0         & H_t    & ~           & ~   \\
        ~           & H_t^\dagger & H_0    & \ddots      & ~   \\
        ~           & ~           & \ddots & \ddots      & H_t \\
        ~           & ~           & ~      & H_t^\dagger & H_0         
    \end{pmatrix},
\end{align}
i.e.\ it only allows for hoppings between nearest neighbors (perpendicular to the surface of interest). For longer hoppings, the unit cell has to be enlarged correspondingly. 
The algorithm for the surface Green's function has initial values
\begin{align}
    \epsilon^t_0 &= \epsilon^b_0 = \epsilon^c_0 = H_0\\
    \alpha_0 &= H_t\\
    \beta_0 &= H_t^\dagger
\end{align}
and the iterative step is:
\begin{align}
    G &= \br{i\omega-\epsilon^c_i}^{-1}\\
    \epsilon^t_{i+1} &= \epsilon^t_i + \alpha_i G \beta_i\\
    \epsilon^b_{i+1} &= \epsilon^b_i + \beta_i G \alpha_i\\
    \epsilon^c_{i+1} &= \epsilon^c_i + \alpha_i G \beta_i + \beta_i G \alpha_i\\
    \alpha_{i+1} &= \alpha_i G \alpha_i\\
    \beta_{i+1} &= \beta_i G \beta_i
\end{align}
The iteration can be stopped if $\alpha_i$ and $\beta_i$ are sufficiently small. The Green's function of the top ($G^t$), bottom ($G^b$), and center ($G^c$) layer is given by:
\begin{align}
    G^t &= \br{i\omega - \epsilon^t_i}^{-1}\\
    G^b &= \br{i\omega - \epsilon^b_i}^{-1}\\
    G^c &= \br{i\omega - \epsilon^c_i}^{-1}
\end{align}
The numerically efficiency in reaching very large $L$ sets the Green's function method apart from a direct real-space slab approach, the numerical cost of which strongly increases with $L$ and which would in addition require disentangling the BC contribution of different layers in the finite slab. On the other hand a real-space slab calculation might in principle accommodate surface lattice reconstructions, even if slab sizes would be moderate, which our present Green's function approach does not.
We use the iterative procedure to access the Green's function of a surface layer of thickness $N$, where $N$ is determined by the choice of unit cell size at the first iteration of the method.
The results presented below are typically for the choice $N=1$ (corresponding to the top primitive unit cell) unless stated otherwise.
$G^S$ has a finite momentum dependent self-energy and only depends on momenta ${\bm k} _\parallel =(k_x,k_y)$ parallel to the surface. Based on it, we define and calculate the associated surface Berry curvature $\Omega^S_z({\bm k}_\parallel)$ by a frequency integral over  $g^S_z(i\omega,{\bm k}_\parallel)$ analogous to Eqs.~(\ref{Eq:BC}) and (\ref{Eq:BC_Green}).

In order to obtain the relevant quantity for transport measurements, the BCD, we use the identity $\frac{\partial f}{\partial k_j} = \frac{\partial f}{\partial \mu}  \frac{\partial G^{-1}}{\partial k_j}$ where $f(\mu - H({\bm k}))$ is the Fermi function, and we rewrite the BCD in terms of (surface) Green's functions as follows:
\begin{align}
	D^S_{a} &= -\int \frac{\text{d}^2{\bm k}_\parallel}{(2\pi)^2} ~\partial_a f(\mu,\vec{k}_\parallel) ~\Omega^S_z({\bm k_\parallel}) \nonumber \\
              &= -\int \frac{\text{d}^2{\bm k}_\parallel}{(2\pi)^2} ~d^S_{a}(\mu,{\bm k}_\parallel)\\
	d^S_{a}({\bm k}_\parallel) &=\int_{-\infty}^\infty \frac{\text{d} \omega}{2\pi}~\text{tr} \Big[ \partial_a \big(G^{S}(\mu,{\bm k}_\parallel)\big)^{-1} ~\partial_\mu g^S_z(i\omega+\mu,{\bm k}_\parallel)\Big] \nonumber
\end{align}
where $D^S_{a}$ is the SBC dipole, $d^S_{a}$ its density in momentum space and we omit their dependence on $\mu$ for brevity.
Since only states at the Fermi surface contribute to the BCD, it is convenient for a numerical integration over the Brillouin zone to broaden the derivative of $g^S_z(i\omega+\mu,\vec{k})$ by approximating it with a finite symmetric difference quotient with width $\Delta\mu$. In order to reduce numerical errors in the $\omega$ integral, we determine the poles of $G^S$ and shift the contour away from the poles if necessary and only calculate the velocity of the bands at the Fermi energy by diagonalizing $G^S$. The integration might require high precision for $G^S$ at the projection of bulk states because $G^S$ diverges at $i\omega=0$ along a whole line. In this case, we checked that leaving a small gap to avoid the line-divergence in the integration contour at the real axis still provides a good approximation.

{\it Density functional calculations for Bi, Rh and HgTe}  ---
Using the Full-Potential-Local-Orbital code, FPLO~\cite{PhysRevB.59.1743}, we constructed symmetry-conserving maximally-projected Wannier functions which render an effective Hamiltonian $H_{\bm k}$ of our 3D system depending on momenta ${\bm k}=(k_x,k_y,k_z)$. This scheme can be regarded as a zeroth-order approximation of the maximally localized approach in projecting a subset of wave functions onto a set of suitably chosen local trial functions with subsequent orthonormalization~\cite{koepernik2021symmetry}. The particular nature of the local orbitals in FPLO make them an efficient set of projectors, since they are constructed to be a “chemical” basis, leading to highly localized Wannier functions, which obey the space group symmetry of the crystal by construction. 

Bismuth crystallizes in the space group R$\bar{3}$m and here we use lattice parameters $a=4.609$\,\AA~and $c=11.975$\,\AA. HgTe has a cubic zincblende structure, space group F$\bar{4}$3m.
In this work we consider both such structure as well as one obtained by applying  $2\%$ tensile strain that causes opening of a gap and stabilizes a topological insulating phase \cite{PhysRevLett.106.126803}.
For this, we first relax the volume by minimization of the total energy obtaining the equilibrium lattice parameter $a_0=6.699$\,\AA. We then enforce the above-mentioned deformation and relax by similar means the lattice parameter $c$. To exploit the remaining symmetries, in this case we work with the space group I$\bar{4}m2$ with lattice parameters $a=b=4.832$\,\AA\, and $c=6.532$\,\AA. 
We work in the generalized gradient approximation and include the spin-orbit coupling in the four-component formalism. We use a linear tetrahedron method for Brillouin zone integrations using a $k$-mesh with $12^3$ subdivisions in the case of Bi and $16^3$ in the case of HgTe. 
For the Wannier models, for Bi we include 6\textit{p} orbitals for every Bi atom and, for HgTe, 6\textit{s} orbitals of Hg and 5\textit{p} of Te. 
For Rh the calculation is based on a Wannier Hamiltonian with 5s and 4d orbitals and to simplify the calculation of the surface Green's function, we kept matrix elements coupling atoms closer than 12\AA.

\section{Acknowledgments}
We thank Inti Sodemann and Jhih-Shih You for fruitful discussions, Ulrike Nitzsche for technical assistance and the DFG for support through the W\"urzburg-Dresden Cluster of Excellence on Complexity and Topology in Quantum Matter, ct.qmat (EXC 2147, project-id 39085490) and through SFB 1143 (project-id 247310070) project A5. J.I.F. would like to thank the support from the Alexander von Humboldt Foundation during the part of his contribution to this work done in Germany and ANPCyT grants PICT 2018/01509 and PICT 2019/00371.

\pagebreak

\pagebreak
.

\pagebreak
.

\end{document}